\documentclass[prc,twocolumn,superscriptaddress,showpacs,amssymb,amsmath,amsfonts,aps]{revtex4}
\usepackage{graphicx}
\usepackage{dcolumn}
\usepackage{epsfig}

\begin{document}
\title{Electroexcitation of the 
$\Delta(1232)\frac{3}{2}^+$ and 
$\Delta(1600)\frac{3}{2}^+$ 
in a light-front relativistic quark model \\}

\newcommand*{\JLAB }{ Thomas Jefferson National Accelerator Facility, 
Newport News, Virginia 23606, USA}
\affiliation{\JLAB }
\newcommand*{\YEREVAN }{ Yerevan Physics Institute, 375036 Yerevan, 
Armenia}
\affiliation{\YEREVAN }
\author{I.G.~Aznauryan}
     \affiliation{\JLAB}
     \affiliation{\YEREVAN}
\author{V.D.~Burkert}
     \affiliation{\JLAB}
\begin{abstract}
{
The magnetic-dipole form factor and the ratios $R_{EM}$ and $R_{SM}$
for the $\gamma^* N\rightarrow \Delta(1232)\frac{3}{2}^+$ transition 
are predicted
within light-front relativistic quark model 
up to photon virtuality $Q^2= 12~$GeV$^2$.
We also predict the helicity amplitudes of the 
$\gamma^* N\rightarrow \Delta(1600)\frac{3}{2}^+$ transition 
assuming the $\Delta(1600)\frac{3}{2}^+$ 
is the first
radial excitation of the ground state $\Delta(1232)\frac{3}{2}^+$. 
}
\end{abstract}
\pacs{ 12.39.Ki, 13.40.Gp, 13.40.Hq, 14.20.Gk}
\maketitle

\section{Introduction}
\label{intro}
One of the longstanding and intriguing problems of hadron physics 
is the identification of the states that can be assigned
as the first  radial excitations of the nucleon and $\Delta(1232)\frac{3}{2}^+$.
It is well recognized 
that the crucial role in the identification
of the Roper resonance N(1440)$\frac{1}{2}^+$ as 
a predominantly first radial excitation of the three-quark
($3q$) ground state belongs to the measurements 
by the CLAS collaboration \cite{Park,Joo,Joo1,Joo2,Egiyan,Biselli} 
that resulted in the determination
of the electrocouplings 
of this resonance
with the proton in a wide range of 
$Q^2=0.3-4.2~$GeV$^2$.
Comparison of the $\gamma^* p\rightarrow $N(1440)$\frac{1}{2}^+$
transition amplitudes extracted from these data \cite{Aznauryan2008,Aznauryan2009}
with the predictions of 
the LF relativistic quark models (LF RQM) \cite{Capstick,Aznauryan2007}
provided strong evidence for the N(1440)$\frac{1}{2}^+$
as a member of the multiplet $[56,0^+]_r$, with additional
non-3-quark contributions needed to describe the low $Q^2$ behavior
of the amplitudes. 

Our goal in this paper is computation of 
the $\gamma^* N\rightarrow \Delta(1600)\frac{3}{2}^+$ 
transition amplitudes in the LF RQM.
Comparison of the results obtained in the quark model 
with the amplitudes 
that are expected to be extracted from experimental data
will provide important test for the commonly expected asignment
of the  $\Delta(1600)\frac{3}{2}^+$ 
as the first  radial excitation of the $\Delta(1232)\frac{3}{2}^+$.
Very recently, the CLAS data on the differential cross sections 
of exclusive process $e p \rightarrow e \pi^+ n$ were reported
in the range of $Q^2 = 1.8 - 4~$GeV$^2$, and the invariant
mass range of the $\pi^+ n$ final state $W = 1.6 - 2.0$~GeV \cite{Park1}.
These data combined with the earlier CLAS data
\cite{Park} on the cross sections and longitudinally polarized beam asymmetries 
for this reaction in the lower mass range $W =1.15- 1.69~$GeV and 
at close values of $Q^2$ allowed the extraction of
the electroexcitation amplitudes of the resonances
$N(1675)\frac{5}{2}^-$,
$N(1680)\frac{5}{2}^+$, and $N(1710)\frac{1}{2}^+$
in the third resonance region.
The isotopic pairs of the resonances from this region:
$\Delta(1600)\frac{3}{2}^+$ and $N(1720)\frac{3}{2}^+$,
$\Delta(1620)\frac{1}{2}^-$ and $N(1650)\frac{1}{2}^-$, and
$\Delta(1700)\frac{3}{2}^-$ and $N(1700)\frac{3}{2}^-$,
could not be separated from each other using data from a single 
isospin channel. 
Currently new data are in preparation by the CLAS collaboration  
for the $ep \to ep\pi^0$ process in the same kinematics region as the data 
in the $ep \to e n \pi^+$ channel \cite{Park,Park1}, as well as at lower $Q^2$.
The two-channel analysis will allow the extraction of
the electroexcitation amplitudes of all resonances 
from the third resonance region including 
the $\Delta(1600)\frac{3}{2}^+$.

The approach we use is based on the LF dynamics and is formulated 
in Refs. \cite{Terentiev1,Aznauryan1982}.
In numerous applications (see Refs. \cite{Aznauryan2007,Aznauryan2012}
and references therein), this approach  
was utilized for the investigation of nucleon form factors and electroexcitation
of nucleon resonances. 

In this work we study the electroexcitation of 
the $\Delta(1600)\frac{3}{2}^+$ in parallel with that of 
the $\Delta(1232)\frac{3}{2}^+$, where we complement the results
obtained earlier in Ref. \cite{Aznauryan2012}
by computing  all three form factors that describe
the transition $\gamma^* N\rightarrow \Delta(1232)\frac{3}{2}^+$. 
In Refs. \cite{Aznauryan85,Keister} it was shown that
there are difficulties in the utilization of the LF 
approaches for hadrons with spin $J\geq 1$.
In the approach of Ref. \cite{Aznauryan1982}, 
these difficulties limit the number of transition amplitudes
that can be investigated for 
the $\Delta(1232)\frac{3}{2}^+$ and $\Delta(1600)\frac{3}{2}^+$. 
Reliable results can be obtained
only for two of the three transition form factors. They are based on the utilization 
of longitudinal components of the electromagnetic current $J_{em}^{0,z}$.
For the  $\Delta(1232)\frac{3}{2}^+$, 
the results obtained for two transition form factors
have been presented in Ref. \cite{Aznauryan2012}.
In the present work, we complement these results
by calculating the third transition form factor
using $J_{em}^{x}+iJ_{em}^{y}$. 
As was shown in Ref. \cite{Aznauryan1982},
these results are less reliable, as  
the matrix elements of transverse components
of the electromagnetic current can contain contributions
that violate impulse approximation, i.e. contributions of diagrams
containing vertices like $\gamma^*\rightarrow q{\bar q}$.
Similar problem exists in the LF RQM of Refs. \cite{Keister,Capstick},
where the requirement of rotational covariance
can not be satisfied without introducing  
two- and three-body current operators. For this reason,
the results for the electroexcitation amplitudes
for the resonances with spins $J=\frac{3}{2}$
are presented in Ref. \cite{Capstick} along with curves
which show the uncertainty that can be caused
by the violation of the rotational covariance.
When presenting our results we also demonstrate the uncertainty
that can arise due to the inclusion of the transverse components
of the electromagnetic current.

An important aspect in the comparison of the transition amplitudes
obtained in theoretical approaches with the amplitudes
extracted from experimental data is their
sign (see, for example, Ref. \cite{Aznauryan_review}).  The results on the 
$\gamma^* N\rightarrow N^*$ transition amplitudes extracted from
experimental data contain an additional sign related to the
vertex of the resonance coupling to the final state hadrons.
In the electroproduction of pions on nucleons this is the relative sign
between the $\pi N N^*$ and $\pi N N$ vertices.
For the Roper resonance, this sign was found in Refs. \cite{Capstick}
and \cite{Aznauryan2007} using, respectively, the ${}^3 P_0$ model
and the approach based on PCAC in the way suggested
in Ref. \cite{Gilman1974}. The results obtained in both approaches
are consistent with each other.  
In Sec. \ref{delta_amplitudes}, we determine the relative signs of the vertices
$\pi N N$, $\pi N \Delta(1232)$, and $\pi N \Delta(1600)$
using the approach based on PCAC.

Our goals and the ranges of $Q^2$, where we make
predictions, for the resonances $\Delta(1232)\frac{3}{2}^+$
and $\Delta(1600)\frac{3}{2}^+$ are different.
For the $\Delta(1600)\frac{3}{2}^+$, we make predictions
that are of interest to reveal the nature of this
resonance using the existing and future CLAS data at $Q^2<4~$GeV$^2$.
For the $\Delta(1232)\frac{3}{2}^+$, our goal is to make
predictions up to $12~$GeV$^2$. These results
will be important for the interpretation of future data on
$\gamma^*p\rightarrow \Delta(1232)\frac{3}{2}^+$ that are expected
with the Jefferson Lab $12~$GeV upgrade.

In Sec. \ref{delta_amplitudes} we present the LF RQM formalism to compute
the $\gamma^* N\rightarrow \Delta$ transition amplitudes.
The results for both resonances are presented and discussed
in Sec. \ref{results} and summarized in Sec. \ref{summary}. 

\section{The $\gamma^* N\rightarrow \Delta$ transition amplitudes
in LF RQM}
\label{delta_amplitudes}
The $\gamma^* N\rightarrow \Delta(1232)\frac{3}{2}^+$ 
and $\gamma^* N\rightarrow \Delta(1600)\frac{3}{2}^+$ amplitudes
have been evaluated within the approach
of Ref. \cite{Aznauryan1982} where the 
LF RQM
is formulated in the infinite momentum frame (IMF).
The IMF is chosen in such a way, that the initial hadron moves
along the $z$-axis with the momentum ${\rm P}\rightarrow \infty$,
the virtual photon momentum is
${\rm k}^{\mu}=\left(
\frac {M^2-m^2-\mathbf{Q}^2_{\perp}}{4{\rm P}},
\mathbf{Q}_{\perp}, 
-\frac {M^2-m^2-\mathbf{Q}^2_{\perp}}{4{\rm P}}\right)$,
the final hadron momentum is
${\rm P'=P+k}$, and ${\rm Q^2\equiv -k^2}=\mathbf{Q}_{\perp}^2$; 
$m$ and $M$
are masses of the nucleon and $\Delta$, respectively. 
In this frame, the matrix elements of the electromagnetic current
for the $\gamma^* N\rightarrow \Delta$ transition
have the form:
\begin{eqnarray}
&& <\Delta,S'_z|J_{em}^{\mu}|N,S_z>|_
{{\rm P}\rightarrow\infty} \nonumber \\
&&=3eQ_a\int \Psi'^+({\rm p}'_a,{\rm p}'_b,{\rm p}'_c) \Gamma_a^\mu
\Psi({\rm p}_a,{\rm p}_b,{\rm p}_c) d\Gamma,
\label{eq:sec1}
\end{eqnarray}
where $S_z$ and $S'_z$ are the projections of the hadron
spins on the $z$-direction.
In Eq. (\ref{eq:sec1}), it is supposed that 
the photon interacts with quark $a$ (the quarks
in hadrons are denoted by $a,b,c$), 
$Q_a$ 
is the charge of this quark in units of $e$ ($e^2/4\pi=1/137$);
$\Psi$ and $\Psi'$ are wave functions
in the vertices $N(\Delta)\leftrightarrow 3q$;
${\rm p}_i$ and ${\rm p}'_i$ ($i=a,b,c$) are the quark momenta
in IMF; 
$d\Gamma$ is the phase space volume;
$\Gamma_a^\mu$ corresponds to the vertex of the quark interaction
with the photon:
\begin{eqnarray}
&& x_a\Gamma_a^{x}=2{\rm p}_{ax}+{\rm Q}_x+i{\rm Q}_y\sigma_z^{(a)},\\
\label{eq:sec2}
&& x_a\Gamma_a^{y}=2{\rm p}_{ay}+{\rm Q}_y-i{\rm Q}_x\sigma_z^{(a)},\\
\label{eq:sec3}
&& \Gamma_a^{0}=\Gamma_a^z=2{\rm P},
\label{eq:sec4}
\end{eqnarray}
where $x_a$ is the fraction of the initial hadron momentum carried
by the quark.

Let $\mathbf{q}_{i}~(i=a,b,c)$ be the three-momenta
of initial quarks in their c.m.s.:
$\mathbf{q}_{a}+\mathbf{q}_{b}+\mathbf{q}_{c}=0$.
The sets of the quark three-momenta in the IMF and in the c.m.s.
of the quarks are related as follows: 
\begin{equation}
\mathbf{p}_{i}=x_i\mathbf{P}+\mathbf{q}_{i\perp},~
~~\sum\limits_{i} {x_i}=1.
\label{eq:sec5}
\end{equation}

According to results of Ref. \cite{Aznauryan1982},
the wave function $\Psi$ is related
to the wave function in the c.m.s. of quarks 
through Melosh matrices \cite{Melosh}:
\begin{equation}
\Psi=U^+({\rm p}_a)U^+({\rm p}_b)U^+({\rm p}_c)\Psi_{fss}
\Phi(\mathbf{q}_{a},\mathbf{q}_{b},\mathbf{q}_{c}).
\label{eq:sec6}
\end{equation}
Here we have separated the flavor-spin-space ($\Psi_{fss}$)
and spatial ($\Phi$)
parts of the c.m.s. wave function. 
The Melosh matrices are 
\begin{equation}
U({\rm p}_i)=\frac{m_q+M_0x_i+i\epsilon _{lm}\sigma_l {\rm q}_{im}}
{\sqrt{(m_q+M_0x_i)^2+\mathbf{q}_{i\perp}^2}},
\label{eq:sec7}
\end{equation}
where $m_q$ is the quark mass and
$M_0$ is invariant mass
of the system of initial quarks:
\begin{equation}
M_0^2=\left(\sum\limits_{i} {{\rm p}_i}\right)^2=
\sum\limits_{i} {\frac{\mathbf{q}_{i\perp}^2+m_q^2}{x_i}}.
\label{eq:sec8}
\end{equation}
In the c.m.s. of quarks:
\begin{equation}
M_0=\sum\limits_{i} 
{\omega_i},~~~\omega_i=\sqrt{m_q^2+\mathbf{q}_{i}^2},
~~~{\rm q}_{iz}+\omega_i=M_0x_i.
\label{eq:sec9}
\end{equation}

For the final state quarks, 
the quantities defined by Eqs. (\ref{eq:sec5}-\ref{eq:sec9}) 
are expressed through ${\rm p}'_i$, $\mathbf{q}'_{i}$,
and $M'_0$.
The phase space volume in Eq. (\ref{eq:sec1})
has the form:
\begin{equation}
d\Gamma=(2\pi)^{-6}\frac
{d\mathbf{q}_{b\perp}d\mathbf{q}_{c\perp}dx_b dx_c}
{4x_ax_bx_c}.
\label{eq:sec10}
\end{equation}

To study sensitivity to the form of the quark wave function,
we employed two forms of the spatial wave function:

\begin{eqnarray} 
&&\Phi_{N(\Delta)}^{(1)}=N_{N(\Delta)}^{(1)} exp(-M_0^2/6\alpha_1^2),
\label{eq:sec11}
\\
&&\Phi_{\Delta_r}^{(1)}=N_{\Delta_r}^{(1)}(\beta_1^2-M_0^2) exp(-M_0^2/6\alpha_1^2)
\label{eq:sec12}
\end{eqnarray}

and

\begin{eqnarray} 
&&\Phi_{N(\Delta)}^{(2)}=N_{N(\Delta)}^{(2)}
exp\left[-({\bf{q}}_a^2+{\bf{q}}_b^2+{\bf{q}}_c^2)/2\alpha_2^2\right],
\label{eq:sec13}
\\
&&\Phi_{\Delta_r}^{(2)}=N_{\Delta_r}^{(2)}
\left[\beta_2^2-({\bf{q}}_a^2+{\bf{q}}_b^2+{\bf{q}}_c^2)\right] 
\nonumber \\
&&~~~~~~~~~~~~exp\left[-({\bf{q}}_a^2+{\bf{q}}_b^2+{\bf{q}}_c^2)/2\alpha_2^2\right],
\label{eq:sec14}
\end{eqnarray}
that were used, respectively, in Refs. 
\cite{Terentiev1,Aznauryan1982}
and \cite{Capstick}. 
The parameters $N$ and $\beta$ are determined by the conditions:
\begin{equation}
\int \Phi^2_{N(\Delta,\Delta_r)} d\Gamma =1,~~~
\int \Phi_{N(\Delta)} \Phi_{\Delta_r} d\Gamma =0.
\label{eq:sec15}
\end{equation}

To distinguish between ground state $\Delta(1232)$ and the
$\Delta(1600)$, considered as the member of the multiplet $[56,0^+]_r$, 
we have used in Eqs. 
(\ref{eq:sec11}-\ref{eq:sec15}) notations $\Delta$ and $\Delta_r$.

Other parameters of the model, namely, the quark mass $m_q$ and 
the oscillator parameter $\alpha$, were found in Ref. \cite{Aznauryan2012}
from the description of nucleon form factors up to $Q^2=16~$GeV$^2$:
\begin{eqnarray}
&&\alpha_1=0.37~{\rm GeV},~~~\alpha_2=0.41~{\rm GeV},
\label{eq:nuc1}\\
&& m_q^{(1)}(Q^2)=\frac{0.22{\rm GeV}}{1+Q^2/56{\rm GeV}^2},
\label{eq:nuc2}\\
&& m_q^{(2)}(Q^2)=\frac{0.22{\rm GeV}}{1+Q^2/18{\rm GeV}^2}.
\label{eq:nuc3}
\end{eqnarray} 

The $Q^2$-dependence of the constituent quark mass
(\ref{eq:nuc2},\ref{eq:nuc3})
is in qualitative agreement with the QCD lattice calculations
and Dyson-Schwinger equations
\cite{Bowman,Bhagwat1,Bhagwat2}, where
the running quark mass is generated dynamically.
The parameters (\ref{eq:nuc1}) and the parameterizations 
(\ref{eq:nuc2},\ref{eq:nuc3})
have been used in the present calculations.
For both resonances, the results for  
the transition amplitudes
obtained with the wave functions
(\ref{eq:sec11},\ref{eq:sec12}) and (\ref{eq:sec13},\ref{eq:sec14}) turned out very close
to each other.

Electroexcitation of the states with $J^P=\frac{3}{2}^+$ on the nucleon
is described by three form factors $G_1(\rm Q^2)$, $G_2(\rm Q^2)$,
and $G_3(\rm Q^2)$, which we define according to
Refs. \cite{Aznauryan_review,Devenish} in the following way:

\begin{eqnarray}
&<\Delta,J^P=\frac{3}{2}^+|J_{em}^{\mu}|N>\equiv e\bar{u}
_{\nu}(\rm P')\gamma_5\Gamma^{\nu\mu}u(\rm P),
\label{eq:sec16}\\
&\Gamma^{\nu\mu}(\rm Q^2)= G_1{\cal H}_1^{\nu\mu}+
G_2{\cal H}_2^{\nu\mu}+G_3{\cal H}_3^{\nu\mu},
\label{eq:sec17}\\
&{\cal H}_1^{\nu\mu}=\rm k\hspace{-1.8mm}\slash
g^{\nu\mu}-\rm k^{\nu}\gamma^{\mu},
\label{eq:sec18}\\
&{\cal H}_2^{\nu\mu}=\rm k^{\nu}P'^{\mu}-(kP')g^{\nu\mu},
\label{eq:sec19}\\
&{\cal H}_3^{\nu\mu}=\rm k^{\nu}k^{\mu}-k^2g^{\nu\mu},
\label{eq:sec20}
\end{eqnarray}
where $u(\rm P)$ and $u_{\nu}(\rm P')$ are, respectively, the
Dirac and Rarita-Schwinger spinors. These form factors have been
found through the matrix elements (\ref{eq:sec1}) using
the relations:

\begin{eqnarray}
&& \frac{1}{2{\rm P}}<\Delta,\frac{3}{2}|J_{em}^{0,z}|N,\frac{1}{2}>|_
{{\rm P}\rightarrow\infty}=\nonumber \\
&& -\frac{{\rm Q}}{\sqrt{2}}
\left[G_1({\rm Q}^2)+\frac{M-m}{2}G_2({\rm Q}^2)\right],
\label{eq:sec21}
\\
&& \frac{1}{2{\rm P}}<\Delta,\frac{3}{2}|J_{em}^{0,z}|N,-\frac{1}{2}>
|_{{\rm P}\rightarrow\infty}=
\nonumber
\\
&&~~~~~~~~~~~~~~~~\frac{{\rm Q}^2}{2\sqrt{2}}G_2({\rm Q}^2),
\label{eq:sec22}
\\
&& <\Delta,\frac{3}{2}|J_{em}^{x}+iJ_{em}^{y}|N,-\frac{1}{2}>
|_{{\rm P}\rightarrow\infty}=
\nonumber
\\
&&~~~~~~~~~~~~~~~~\frac{{\rm Q}^3}{\sqrt{2}}G_3({\rm Q}^2).
\label{eq:sec23}
\end{eqnarray}

The relations between form factors 
$G_1(\rm Q^2)$, $G_2(\rm Q^2)$, and $G_3(\rm Q^2)$
and the $\gamma^* N\rightarrow \Delta(\frac{3}{2}^+)$
helicity amplitudes and the Jones-Scadron 
form factors $G_M(Q^2)$, $G_E(Q^2)$, and $G_C(Q^2)$ \cite{J-S}
are given in the Appendix.

In the approach based on PCAC, the
relative signs of the
$\pi N N$, $\pi N \Delta(1232)$, and $\pi N \Delta(1600)$
vertices are determined according to 
Refs. \cite{Aznauryan85,Aznauryan2007}
by the relative signs of the following  expressions:

\begin{equation}
I_{NA}\equiv \int\frac
{(m_q+M_0x_a)^2-\mathbf{q}_{a\perp}^2}
{(m_q+M_0x_a)^2+\mathbf{q}_{a\perp}^2}
\Phi_N(M_0^2)
\Phi_A(M_0^2)d\Gamma,
\label{eq:sec24}
\end{equation}
where $A$ denotes the states $N$, $\Delta(1232)$, and $\Delta(1600)$.
Numerical calculation of $I_{NN}$, $I_{N\Delta(1232)}$, and
$I_{N\Delta(1600)}$ with the wave functions (\ref{eq:sec11}-\ref{eq:sec14})
gives positive relative signs for the
$\pi N N$, $\pi N \Delta(1232)$, and $\pi N \Delta(1600)$
vertices.

\section{Results}
\label{results}

\subsection{The $\Delta(1232)\frac{3}{2}^+$ resonance}
\label{delta}

\begin{figure}[htp]
\begin{center}
\includegraphics[width=7.6cm]{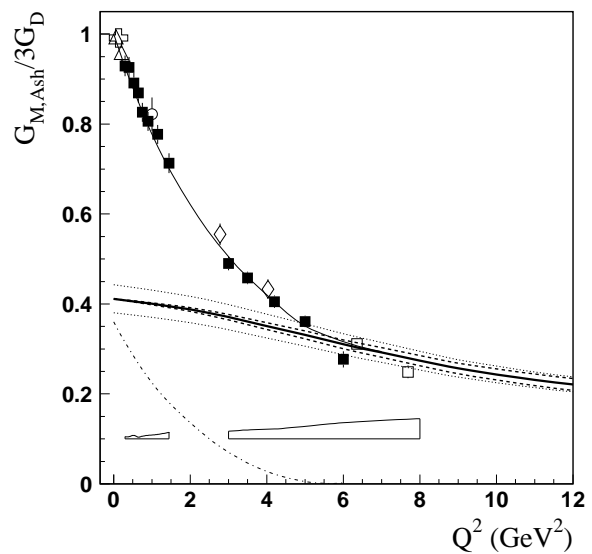}
\vspace{-0.1cm}
\caption{\small
The form factor $G_{M,Ash}(Q^2)$
for the  $\gamma^* p \rightarrow~\Delta(1232)\frac{3}{2}^+$
transition relative to $3G_D$: $G_D(Q^2)=1/(1+\frac{Q^2}{0.71GeV^2})$.
The full boxes are the CLAS data extracted
in the analysis of Ref. \cite{Aznauryan2009},
the open boxes correspond to the data from Ref. \cite{Vilano}.
The bands show the model uncertainties of these data
\cite{Aznauryan2009,Aznauryan_review}.
The thin solid curve is the result of the global analysis of the
Mainz group \cite{Mainz}.
The results from other experiments are: open triangles
\cite{Stave2006,Sparveris2007,Stave2008},
open cross \cite{Mertz,Kunz,Sparveris2005}, open rhombuses \cite{Frolov},
and open circle \cite {KELLY1,KELLY2}.
The thick solid curve presents our results;
the dashed curves are our results corresponding to
$\pm 50\%$ deviation of
$G_3(Q^2)$ 
from the values obtained using the relation   
(\ref{eq:sec23});
the dotted curves show the uncertainty of our
results (given by the solid curve) that is caused by
the uncertainty of $c_{\Delta}$ 
(\ref{eq:sec28.1}).
The dashed-dotted curve is meson-baryon contribution from Refs. \cite{Lee,Lee1}.
\label{gm}}
\end{center}
\end{figure}

\begin{figure}[htp]
\begin{center}
\includegraphics[width=7.6cm]{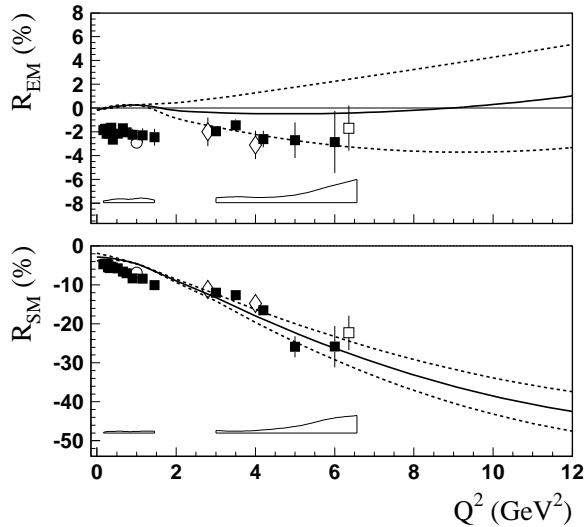}
\vspace{-0.1cm}
\caption{\small
The ratios $R_{EM},~R_{SM}$
for the  $\gamma^* p \rightarrow~\Delta(1232)\frac{3}{2}^+$
transition. The legend for experimental data and thick
solid and dashed curves is as for Fig. \ref{gm}.
\label{ratios}}
\end{center}
\end{figure}

We present the results for the $\Delta(1232)\frac{3}{2}^+$
in terms of the $\gamma^* p \rightarrow~\Delta(1232)\frac{3}{2}^+$
magnetic-dipole transition form factor in the Ash convention
\cite{Ash} (Fig. \ref{gm}) and the ratios 
$R_{EM}\equiv ImE^{3/2}_{1+}/ImM^{3/2}_{1+}$ and
$R_{SM}\equiv ImS^{3/2}_{1+}/ImM^{3/2}_{1+}$ (Fig. \ref{ratios}).
These observables are commonly used to present the results on the 
$\Delta(1232)\frac{3}{2}^+$ extracted from experimental data on
the electroproduction of pions on nucleons.
The $\gamma^* p \rightarrow~\Delta(1232)\frac{3}{2}^+$
magnetic-dipole form factor in the Ash convention is related
to the Jones-Scadron form factor 
defined in the Appendix  as follows:

\begin{equation}
G_{M,Ash}(Q^2)=\frac{G_{M}(Q^2)}{\sqrt{1+\frac{Q^2}{(M+m)^2}}}.
\label{eq:sec25}
\end{equation}

The ratios
$R_{EM}$ and $R_{SM}$ are related to the Jones-Scadron
form factors by: 

\begin{equation}
R_{EM}=-\frac{G_E}{G_M},~~~
R_{SM}=-\frac{G_C}{G_M}\frac{K}{2m},
\label{eq:sec26}
\end{equation}
where $K$ is the virtual photon 3-momentum in the 
c.m.s. of the reaction $\gamma^* N\rightarrow \pi N$:

\begin{equation}
K\equiv \frac{\sqrt{Q_+Q_-}}{2M},
~~~Q_{\pm}\equiv (M \pm m)^2+{\rm Q}^2.
\label{eq:sec27}
\end{equation}

As it was mentioned in the Introduction, in the approach
that we utilize \cite{Aznauryan1982}, 
the results are reliable that are obtained
through longitudinal components of the electromagnetic current $J_{em}^{0,z}$,
i.e. the results for the form factors $G_1(Q^2)$ and $G_2(Q^2)$
(\ref{eq:sec21},\ref{eq:sec22}).
These results have been
presented and discussed in Ref. \cite{Aznauryan2012}.
In this paper, we complement the results
for $G_1(Q^2)$ and $G_2(Q^2)$
by calculating the third transition form factor $G_3(Q^2)$
using $J_{em}^{x}+iJ_{em}^{y}$ 
(\ref{eq:sec23}). This allows us to present the
predictions in a more convenient way in terms of 
$G_{M,Ash}$ and $R_{EM}$ and $R_{SM}$. In order to demonstrate
the uncertainty that can arise due to inclusion of 
the transverse components of the electromagnetic current,
we also present in Figs. \ref{gm},\ref{ratios} results that
correspond to the values of $G_3(Q^2)$ taken with $\pm 50 \%$
deviation from the values obtained using the relation   
(\ref{eq:sec23}). 

It is known that at relatively small $Q^2$, nearly massless
pions generate pion-loop contributions that significantly
alter three-quark contribution to
$\gamma^* p \rightarrow~\Delta(1232)\frac{3}{2}^+$ 
\cite{Bermuth,Thomas,Faessler1}. 
It is expected that the corresponding
hadronic component, including contributions from other mesons,
will be rapidly losing strength with increasing $Q^2$. 
From the description of the data on
pion electroproduction on proton within dynamical reaction model
\cite{Lee,Lee1}, 
it follows that the contribution associated with the meson-baryon
contribution  to  
$\gamma^* p \rightarrow~\Delta(1232)\frac{3}{2}^+$
(dashed-dotted curve in Fig. \ref{gm})
can be neglected
above $Q^2 = 4~$GeV$^2$. 
Therefore, the weight of the $3q$ contribution
to the 
$\Delta(1232)\frac{3}{2}^+$:
\begin{equation}
|\Delta(1232)>=c_{\Delta}|3q>+...,
\label{eq:sec28}
\end{equation}
was found in Ref. \cite{Aznauryan2012} from the description
of the form factors $G_1(Q^2)$ and $G_2(Q^2)$
at $Q^2 > 4~$GeV$^2$: 
\begin{equation}
c_{\Delta}=0.53\pm 0.04.
\label{eq:sec28.1}
\end{equation}
The uncertainty of $c_{\Delta}$ is caused mainly
by the systematic uncertainties of the data on 
$G_{M,Ash}(Q^2)$ at these $Q^2$.
We have used the value of $c_{\Delta}$ from Eq. (\ref{eq:sec28.1})
to find the three-quark contributions to
$G_{M,Ash}(Q^2)$
and $R_{EM}$ and $R_{SM}$,
that are presented in Figs. \ref{gm},\ref{ratios}. 

From the discussion above, it follows that at  
$Q^2 < 4~$GeV$^2$, meson-baryon contributions 
alter the three-quark contribution to
$\gamma^* p \rightarrow~\Delta(1232)\frac{3}{2}^+$
\cite{Bermuth,Thomas,Faessler1,Lee,Lee1}.
With this, for the magnetic-dipole form factor, these contributions
definitely result in better agreement with experiment. 
Above $4-5~$GeV$^2$, we expect that 
the  $\gamma^* p \rightarrow~\Delta(1232)\frac{3}{2}^+$
transition  
will be determined 
by the three-quark contribution only. Therefore, we consider
our results at these $Q^2$ as predictions
for the  $\gamma^* p \rightarrow~\Delta(1232)\frac{3}{2}^+$
transition amplitudes obtained within nonperturbative
approach.

For the form factor $G_{M,Ash}(Q^2)$, the spread of our
results caused by uncertainties
in the form factor $G_3(Q^2)$ is insignificant,
and we have definite predictions
up to $12~$GeV$^2$. According to these predictions, 
above $5~$GeV$^2$ the behaviour
of the ratio $G_{M,Ash}(Q^2)/G_D(Q^2)$ becomes 
more flat in comparison with that at lower $Q^2$.
The similar $Q^2$-dependence is observed for the proton
magnetic form factor \cite{Jones}. For the Jones-Scadron
magnetic-dipole form factor $G_{M}(Q^2)$ and the proton
magnetic form factor $G_{M,p}(Q^2)$, the $Q^2$-dependences
at $Q^2=5-12~$GeV$^2$ practically coincide.

For the ratios $R_{EM}$ and $R_{SM}$, the spread of predictions
grows from $6$ to $10\%$, when $Q^2$ is increasing from $5$ to
$12~$GeV$^2$. Nevertheless, for the ratio $R_{SM}$ one can definitely
conclude, that according to our predictions $R_{SM}$ will
continue to grow and 
within the $Q^2 = 12$~GeV$^2$ limit will
not reach the value predicted  
in pQCD, i.e. $R_{SM}\rightarrow \rm{const}$ with undefined sign and magnitude. 
On the other hand, in holographic QCD in the large $N_c$ limit
the $R_{SM}$ ratio is predicted at the specific asymptotic value: 
$R_{SM}\rightarrow -100\%$ \cite{Grigoryan:2009pp}. The data
show the correct trend, 
but are projected to reach only 40 to 50\% of that value at  $Q^2 \leq 12$~GeV$^2$.

\subsection{The $\Delta(1600)\frac{3}{2}^+$ resonance}
\label{delta1}

\begin{figure*}
\begin{center}
\includegraphics[width=16.8cm]{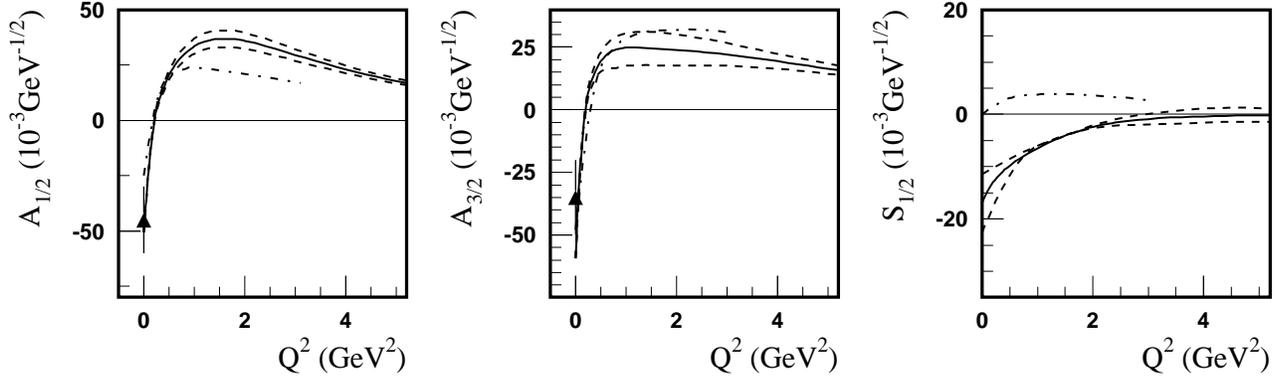}
\vspace{-0.1cm}
\caption{\small
Helicity amplitudes
for the \protect $\gamma^* p \rightarrow~\Delta(1600)\frac{3}{2}^+$
transition.
The full triangles at $Q^2=0$ are the RPP estimates~\cite{Agashe:2014kda}.
The thick solid curve presents our results.
The legend for dashed curves is as for Fig. \ref{gm}.
The dashed-dotted curves present the predictions from Ref. \cite{Capstick}.
\label{hel_ampl}}
\end{center}
\end{figure*}

The results for the resonance  $\Delta(1600)\frac{3}{2}^+$
are presented in Fig. \ref{hel_ampl} in terms of
the $\gamma^* p \rightarrow~\Delta(1600)\frac{3}{2}^+$ helicity amplitudes.
The predictions of the LF RQM approach from Ref. \cite{Capstick}
are also shown. 
The common sign of the
amplitudes 
has been found in our approach and in Ref. 
\cite{Capstick} due to additional computation of the relative signs
 of the $\pi N N$, $\pi N \Delta(1232)$, and $\pi N \Delta(1600)$
vertices using different approaches.

In Section \ref{intro} we have discussed
the difficulties in the utilization of the LF
approach for hadrons with spin $J\geq 1$.
In the approach of  Ref. \cite{Capstick} 
the uncertainty that can be caused by these difficulties
for the  $\Delta(1600)\frac{3}{2}^+$
nearly coincides  
with the longitudinal helicity amplitude
$S_{1/2}$. In our approach these uncertainties are presented
by dashed curves that correspond to the results obtained
with the values of $G_3(Q^2)$ taken with $\pm 50 \%$
deviation from the values obtained using the relation
(\ref{eq:sec23}).
From the presented results we conclude that
independently of uncertainties, both approaches give close
predictions for the transverse helicity amplitudes:
these amplitudes, being negative at $Q^2=0$, change their
signs at $Q^2=0.2-0.3~$GeV$^2$ and become positive.
With this, the values of transverse amplitudes at $Q^2=0$
are in good agreement with the RPP estimates~\cite{Agashe:2014kda}
As in the case of the Roper resonance,
these features will be crucial for conclusions
on the nature of the resonance  $\Delta(1600)\frac{3}{2}^+$
that will be obtained from the comparison with the future
data on the 
$\gamma^* p \rightarrow~\Delta(1600)\frac{3}{2}^+$ helicity amplitudes.

\section{Summary}
\label{summary}
We have employed the LF RQM to evaluate the quark core contributions 
to the transitions $\gamma^* N\rightarrow \Delta(1232)\frac{3}{2}^+$
and $\gamma^* N\rightarrow \Delta(1600)\frac{3}{2}^+$.
Our previous evaluation of the 3-quark 
core contribution to the $\Delta(1232)\frac{3}{2}^+$ 
based on the 
$\gamma^* N\rightarrow \Delta(1232)\frac{3}{2}^+$ 
data up to $Q^2 = 7.5$~GeV$^2$ 
allowed us to make projections into unmeasured territory of $Q^2 \leq 12$~GeV$^2$. 
This region may be covered in upcoming measurements with CLAS12 
at the Jefferson Lab 12 GeV upgrade. 
The projections are made for the magnetic-dipole form factor and electric and scalar 
quadrupole ratios $R_{EM}(Q^2)$ and $R_{SM}(Q^2)$.  
Predictions for the 3 electrocoupling amplitudes are also made 
for the $\Delta(1600)\frac{3}{2}^+$ in the range $Q^2 \leq 5$~GeV$^2$ 
assuming this state is the first radial excitation of 
the $\Delta(1232)\frac{3}{2}^+$. 
The predicted very rapid transition from large negative values 
at the real photon point to large positive values with maxima 
near $Q^2 = 1 - 2$~GeV$^2$, and a slow falloff with $Q^2$ for the two 
transverse amplitudes, should be readily accessible to experimental exploration. 
 
\vspace{0.3cm}

 \section{Acknowledgments}
This work was supported by the US Department of Energy under
contract DE-AC05-06OR23177 and the National Science Foundation
and State Committee of Science of Republic of Armenia,
Grant-13-1C023.

\vspace{0.3cm}

\section{Appendix.
The relations between 
the $\gamma^* N\rightarrow \Delta(\frac{3}{2}^+)$ form factors and 
helicity amplitudes}
\vspace{0.3cm}
\renewcommand\theequation{A\arabic{equation}}
\setcounter{equation} 0

The relations between the form factors $G_1(\rm Q^2)$, $G_2(\rm Q^2)$,
and $G_3(\rm Q^2)$ defined by Eqs. (\ref{eq:sec16}-\ref{eq:sec20}) and   
the $\gamma^* N\rightarrow \Delta(\frac{3}{2}^+)$
helicity amplitudes are following \cite{Aznauryan_review,Devenish}:
 
\begin{equation}
{A}_{1/2}=h_3X,~~{ A}_{3/2}=\sqrt{3}h_2X,
~~{S}_{1/2}=h_1\frac{K}{\sqrt{2}M}X,
\label{eq:ap1}
\end{equation}

where

\begin{eqnarray}
&&h_1({\rm Q}^2)=4MG_1({\rm Q}^2)+4M^2G_2({\rm Q}^2)+\nonumber \\
&&~~~~~~~~~~~~~2(M^2-m^2-{\rm Q}^2)G_3({\rm Q}^2),
\label{eq:ap2}
\\
&&h_2({\rm Q}^2)=-2(M+ m)G_1({\rm Q}^2)-\nonumber \\
&&~~~~(M^2-m^2-{\rm Q}^2)G_2({\rm Q}^2)+
2{\rm Q}^2G_3({\rm Q}^2),
\label{eq:ap3}
\\
&&h_3({\rm Q}^2)=-\frac{2}{M}[{\rm Q}^2+m(
M+m)]G_1({\rm Q}^2)+\nonumber \\
&&~~~(M^2-m^2-{\rm Q}^2)G_2({\rm Q}^2)-2{\rm Q}^2G_3({\rm Q}^2),
\label{eq:ap4}
\\
&&X\equiv e\sqrt{\frac{Q_-}
{48m(M^2-m^2)}}.
\label{eq:ap5}
\end{eqnarray}

The Jones-Scadron 
form factors $G_M(Q^2)$, $G_E(Q^2)$, and $G_C(Q^2)$ \cite{J-S}
are defined by:

\begin{eqnarray}
&&G_M({\rm Q}^2)=-Y(\sqrt{3} A_{3/2}+A_{1/2}),
\label{eq:ap6}
\\
&&G_E({\rm Q}^2)=-Y(A_{3/2}/\sqrt{3}-A_{1/2}),
\label{eq:ap7}
\\
&&G_C({\rm Q}^2)=2\sqrt{2}\frac{M}{K}YS_{1/2},
\label{eq:ap8}
\\
&&Y\equiv \frac{m}{e(M+m)}\sqrt{\frac{2m(M^2-m^2)}{Q_-}}.
\label{eq:ap9}
\end{eqnarray}


\begin{thebibliography}{999}


\bibitem{Joo} K. Joo et al., CLAS Collaboration,
Phys. Rev. Lett. $\bf{88}$, 122001 (2002).

\bibitem{Joo1} K. Joo et al., CLAS Collaboration,
Phys. Rev. C $\bf{68}$, 032201 (2003).

\bibitem{Joo2} K. Joo et al., CLAS Collaboration,
Phys. Rev. C $\bf{70}$, 042201 (2004).

\bibitem{Egiyan} H. Egiyan et al., CLAS Collaboration,
Phys. Rev. C $\bf{73}$, 025204 (2006).

\bibitem{Biselli} A. Biselli et al., CLAS Collaboration,
Phys. Rev. C $\bf{78}$, 045204 (2008).

\bibitem{Park} K. Park et al., CLAS Collaboration,
Phys. Rev. C $\bf{77}$, 015208 (2008).

\bibitem{Aznauryan2008} I. G. Aznauryan et al., CLAS Collaboration,
Phys. Rev. C $\bf{78}$, 045209 (2008).

\bibitem{Aznauryan2009} I. G. Aznauryan et al., CLAS Collaboration,
Phys. Rev. C $\bf{80}$, 055203 (2009).

\bibitem{Capstick} S. Capstick and B. D. Keister, Phys. Rev. D
{\bf 51}, 3598 (1995).

\bibitem{Aznauryan2007} I. G. Aznauryan, 
Phys. Rev. C {\bf 76}, 025212 (2007).

\bibitem{Park1} K. Park et al., CLAS Collaboration,
Phys. Rev. C $\bf{91}$, 045203 (2015).

\bibitem{Terentiev1} L. A. Kondratyuk and M. V. Terent'ev, 
Yad. Fiz., {\bf 31}, 1087 (1980).

\bibitem{Aznauryan1982} I. G. Aznauryan, A. S. Bagdasaryan,  
and N. L. Ter-Isaakyan, 
Phys. Lett. B {\bf 112}, 393 (1982);
Yad. Fiz. {\bf 36}, 1278 (1982).

\bibitem{Aznauryan2012} I. G. Aznauryan and V. D. Burkert, 
Phys. Rev. C {\bf 85}, 055202 (2012).

\bibitem{Aznauryan85} I. G. Aznauryan and A. S. Bagdasaryan,
Sov. J. Nucl. Phys. {\bf 41}, 158 (1985).

\bibitem{Keister} B. D. Keister, Phys. Rev.
{\bf D49}, 1500 (1994).

\bibitem{Aznauryan_review} I. G. Aznauryan, V. D. Burkert, 
Prog. Part. Nucl. Phys. {\bf 67}, 1 (2012), arXiv:1109.1720, 2011.

\bibitem{Gilman1974} F. J. Gilman, M. Kugler, and S. Meshkov, 
Phys. Rev. D {\bf 9}, 715 (1974).

\bibitem{Melosh} H. J. Melosh, Phys. Rev. D
{\bf 9}, 1095 (1974).

\bibitem{Bowman} P. O. Bowman et al.,
Phys. Rev. D {\bf 71}, 054507 (2005).

\bibitem{Bhagwat1} M.S. Bhagwat et al.,
Phys. Rev. C {\bf 68}, 015203 (2003).

\bibitem{Bhagwat2} M.S. Bhagwat and P. C. Tandy,
AIP. Conf. Proc. {\bf 842}, 225 (2006).

\bibitem{Devenish} R.C.E. Devenish, T.S. Eisenschitz,
and J.G. K\"orner,
Phys. Rev. D {\bf 14}, 3063 (1976).

\bibitem{J-S} H. F. Jones and M. D. Scadron,
Ann. Phys. {\bf 81}, 1 (1973).

\bibitem{Ash} W.W. Ash,
Phys. Lett. B {\bf 24},165 (1967).

\bibitem{Vilano} A.N. Villano et al.,
Phys. Rev. C {\bf 80}, 035203 (2009).

\bibitem{Mainz} D. Drechsel, S. Kamalov, and L. Tiator,
Eur. Phys. J. A {\bf 34}, 69 (2007).

\bibitem{Stave2006} S. Stave et al.,
Eur. Phys. J. A {\bf 30}, 471 (2006).

\bibitem{Sparveris2007} N.F. Sparveris et al.,
Phys. Lett. B {\bf 651}, 102 (2007).

\bibitem{Stave2008} S. Stave et al.,
Phys. Rev. C {\bf 78}, 025209 (2008).

\bibitem{Mertz} C. Mertz et al.,
Phys. Rev. Lett. {\bf 86}, 2963 (2001).

\bibitem{Kunz} C. Kunz et al.,
Phys. Lett. B {\bf 564}, 21 (2003).

\bibitem{Sparveris2005} N.F. Sparveris et al.,
Phys. Rev. Lett. {\bf 94}, 022003 (2005).


\bibitem{Frolov} V.V. Frolov et al.,
Phys. Rev. Lett. {\bf 82}, 45 (1999).

\bibitem{KELLY1} J. J. Kelly et al.,
Phys. Rev. Lett. {\bf 95}, 102001 (2005).

\bibitem{KELLY2} J. J. Kelly et al.,
Phys. Rev. C {\bf 75}, 025201 (2007).

\bibitem{Lee} T. Sato and T.-S. H. Lee,
Phys. Rev. C {\bf 63}, 055201 (2001).

\bibitem{Lee1} V. D. Burkert and T.-S. H. Lee,
Int. J. Mod. Phys. E {\bf 13}, 1035 (2004).

\bibitem{Bermuth} K. Bermuth, D. Drechsel, L. Tiator,
and J.B. Seaborn
Phys. Rev. D {\bf 37}, 89 (1988).

\bibitem{Thomas} D.H. Lu, A.W. Thomas, and A.G. Williams,
Phys. Rev. C {\bf 55}, 3108 (1997).

\bibitem{Faessler1} A. Faessler, T. Gutsche, B.R. Holstein et al.,
Phys. Rev. D {\bf 74}, 074010 (2006).

\bibitem{Jones} M. K. Jones et al.,
Phys. Rev. Lett. {\bf 84}, 1398 (2000).

\bibitem{Grigoryan:2009pp}
  H. R. Grigoryan, T.-S. H. Lee and H. U. Yee,
  Phys. Rev. D {\bf 80}, 055006 (2009).

\bibitem{Agashe:2014kda} 
  K.~A.~Olive {\it et al.}  [Particle Data Group Collaboration],
  Chin.\ Phys.\ C {\bf 38}, 090001 (2014).

\end{thebibliography}
\end{document}